\documentclass[sigconf]{acmart}

\AtBeginDocument{%
  }

\usepackage{algorithmic}
\usepackage[ruled,linesnumbered]{algorithm2e}
\usepackage{array}
\usepackage{amsfonts}
\usepackage{amssymb}
\usepackage{amsmath}
\usepackage{bm}
\usepackage{booktabs}
\usepackage{color}
\usepackage{epstopdf}
\usepackage{enumitem}
\usepackage{footmisc}
\usepackage{graphicx}
\usepackage{hyperref}
\usepackage{listings}
\usepackage{multirow}
\usepackage{mathrsfs}
\usepackage{pifont}
\usepackage{subfigure}
\usepackage{setspace}
\usepackage{soul}
\usepackage{textcomp}
\usepackage{url}
\usepackage{xcolor}
\usepackage{xspace}
\usepackage[normalem]{ulem}
\usepackage[np, autolanguage]{numprint}
\usepackage[skip=0pt]{caption}
\usepackage{marvosym}

\usepackage{bbding}

\newcommand{\paratitle}[1]{\vspace{1.5ex}\noindent\textbf{#1}}
\newcommand{\ie}{\emph{i.e.,}\xspace}

\newcommand{\eg}{\emph{e.g.,}\xspace}

\newcommand{\wrt}{w.r.t.\xspace}
\newcommand{\ignore}[1]{}

\newcommand{\modelname}{MDL}

\begin{document}

\title{MDL: A Unified Multi-Distribution Learner in Large-scale Industrial Recommendation through Tokenization}

\author{Shanlei Mu}
\email{mushanlei@bytedance.com}
\authornote{Equal contribution.}
\affiliation{
    \institution{ByteDance Search}
    \city{Beijing}
    \country{China}
}

\author{Yuchen Jiang}
\email{jiangyuchen.jyc@bytedance.com}
\authornotemark[1]
\affiliation{
    \institution{ByteDance AML}
    \city{Beijing}
    \country{China}
}

\author{Shikang Wu}
\email{wushikang@bytedance.com}
\authornotemark[1]
\affiliation{
    \institution{ByteDance Search}
    \city{Beijing}
    \country{China}
}

\author{Shiyong Hong}
\email{hongshiyong.66@bytedance.com}
\affiliation{
    \institution{ByteDance Search}
    \city{Beijing}
    \country{China}
}

\author{Tianmu Sha}
\email{shatianmu@bytedance.com}
\affiliation{
    \institution{ByteDance Search}
    \city{Beijing}
    \country{China}
}

\author{Junjie Zhang}
\email{zhangjunjie.leo@bytedance.com}
\affiliation{
    \institution{ByteDance Search}
    \city{Beijing}
    \country{China}
}

\author{Jie Zhu}
\email{zhujie.zj@bytedance.com}
\affiliation{
    \institution{ByteDance AML}
    \city{Hangzhou}
    \country{China}
}

\author{Zhe Chen}
\email{chenzhe.john@bytedance.com}
\affiliation{
    \institution{ByteDance AML}
    \city{Beijing}
    \country{China}
}

\author{Zhe Wang}
\authornote{Corresponding author.}
\email{zhewang.tim@gmail.com}
\affiliation{
    \institution{ByteDance Search}
    \city{Beijing}
    \country{China}
}

\author{Jingjian Lin}
\email{linjingjian000@gmail.com}
\affiliation{
    \institution{ByteDance Search}
    \city{Beijing}
    \country{China}
}

\renewcommand{\shortauthors}{Shanlei Mu, et al.}

\begin{abstract}
Industrial recommender systems increasingly adopt multi-scenario learning (MSL) and multi-task learning (MTL) to handle diverse user interactions and contexts, but existing approaches suffer from two critical drawbacks: (1) underutilization of large-scale model parameters due to limited interaction with complex feature modules, and (2) difficulty in jointly modeling scenario and task information in a unified framework.
To address these challenges, we propose a unified \textbf{M}ulti-\textbf{D}istribution \textbf{L}earning (MDL) framework, 
inspired by the "prompting" paradigm in large language models (LLMs).
MDL treats scenario and task information as specialized tokens rather than auxiliary inputs or gating signals.
Specifically, we introduce a unified information tokenization module that transforms features, scenarios, and tasks into a unified tokenized format.
To facilitate deep interaction, we design three synergistic mechanisms: (1) feature token self-attention for rich feature interactions, (2) domain-feature attention for scenario/task-adaptive feature activation, and (3) domain-fused aggregation for joint distribution prediction.
By stacking these interactions, MDL enables scenario and task information to "prompt" and activate the model's vast parameter space in a bottom-up, layer-wise manner.
Extensive experiments on real-world industrial datasets demonstrate that MDL significantly outperforms state-of-the-art MSL and MTL baselines.
Online A/B testing on Douyin Search platform over one month yields +0.0626\% improvement in LT30 and -0.3267\% reduction in change query rate.
MDL has been fully deployed in production, serving hundreds of millions of users daily.
\end{abstract}



\maketitle

\section{Introduction}
\label{sec-intro}
Nowadays, recommender systems play an indispensable role in meeting users’ personalized interests and alleviating the information overload problem.
In large-scale industrial recommendation, the recommender model usually has to deal with multiple scenarios (\eg homepage feeds or banner) and multiple user behaviors (\eg click or like), referred to as multi-scenario learning (MSL)~\cite{msl-benchmark} and multi-task learning (MTL)~\cite{mtl-survey} in recommender system.

In recommendations, multi-scenario learning (MSL)~\cite{PLE,HMoE,STAR,PEPNet,hinet,mmfi,m3oe,ADL} aims to learn data from multiple scenarios within a unified model, and multi-task learning (MTL)~\cite{sharedbottom,MMoE,Cross-Stitch,snr,adatt} aims to optimize the unified model with multiple related objectives.
Technically speaking, the key of MSL and MTL in recommendations lies in the modeling of multi-distribution interrelations, capturing the commonalities as well as discriminating the differences across scenarios and tasks.
Most existing methods usually adopt a \emph{shared-specific} framework, where scenario/task-shared and scenario/task-specific components are combined in the backbone.
For example, one class of methods~\cite{MMoE,HMoE,PLE} is based on the Mixture-of-Experts (MoE) structure, which models different scenarios and tasks by designing both shared experts and scenario/task-specific expert.
Another class of methods~\cite{STAR,PEPNet} is based on the dynamic parameter generation approach, where scenario- or task-specific prior information is used to generate dedicated network parameters (such as MLPs or Gate units), enabling differentiated modeling across scenarios and tasks.

Although existing methods have achieved notable results, there are still several issues that remain to be addressed.
(1) \emph{Limitation in scaling capability}: inspired by the scaling law in large language models (LLMs)~\cite{LLM1,LLM2}, recent industrial recommender systems focus on improving model learning capability by scaling up the model parameters of feature-interaction modules, achieving remarkable results~\cite{RankMixer,KGBRank,OneTrans,wukong,HSTU,MTGR,hyformer}.
Typically, the parameter size of the state-of-the-art models is now hundreds of times larger than that of traditional recommender models, usually exceeding 0.1B parameters~\cite{RankMixer,KGBRank}.
Based on large-parameter recommender models, existing MSL and MTL methods lack effective utilization of complex feature interaction modules.
Because these methods typically only affect certain intermediate modules or shallow output layers, making it difficult to fully leverage the benefits of larger model parameters.
(2) \emph{Difficulty in uniformity modeling}: multi-scenario modeling primarily focuses on differences in input distributions, while multi-task modeling emphasizes differences in label distributions.
Traditional shared-specific structures are often designed differently for these two aspects.
It is challenging to jointly model scenario information and task information in a unified way.

To address the above issues, we further draw inspiration from LLMs to enhance MSL and MTL in recommendations.
LLMs accomplish specific downstream tasks by inputting appropriate prompt information~\cite{gpt4,qwen3}.
As a special type of token, prompt tokens can fully activate the vast latent knowledge of LLMs through the self-attention mechanism.
Similarly, we posit that the scenario and task information can serve as prompts within recommender models like that in LLMs.
By tokenizing scenario and task priors into a unified latent space, we transform them from mere auxiliary inputs into active participants within the core feature-interaction backbone.
Instead of restricting scenario/task information to the "shallows" of the model (e.g., gates or heads), tokenized priors can penetrate deep and multi-layered interaction modules. This allows distribution-specific signals to adaptively modulate the model's massive parameter space, fully harvesting the benefits of the scaling law (\emph{corresponding to the first issue}).
By projecting both input-side scenario contexts and output-side task objectives into a consistent tokenized format, we bridge the gap between MSL and MTL. This enables a holistic and joint modeling framework where scenario-task dependencies are captured through a unified interaction protocol (\emph{corresponding to the second issue}).

To this end, we propose the \textbf{M}ulti-\textbf{D}istribution \textbf{L}earning (\textbf{MDL}) framework, a unified approach for multi-scenario and multi-task learning in large-scale reommender system.
At its core, MDL adopts a "Tokenize-and-Interact" philosophy: it tokenizes feature, scenario, and task information in a unified way, then elaborate effective interactions among these tokens to activate the model capability for MSL and MTL.
Specifically, we first use a \emph{unified information tokenization} module to resolve the representational gap between disparate distributions.
This module applies semantic segmentation and nonlinear transformations to raw input features, projecting them into a discrete set of feature tokens, scenario tokens, and task tokens.
Based on these tokens, we facilitate multi-granular information exchange through three specialized mechanisms.
The feature token self-interaction adopts a self-attention-like mechanism to capture high-order feature dependencies, providing the recommender with fundamental expressive capabilities.
The interaction between feature tokens and scenario/task tokens is achieved through \emph{domain-aware attention}, 
where scenario and task tokens act as "queries" that adaptively activate and aggregate relevant feature information. 
By doing so, the model learns to prioritize different feature sub-spaces based on the specific scenario/task context, effectively handling distribution shifts.
Finally, the interaction between scenario and task tokens is achieved through the \emph{domain-fused module}, which aggregates token information to enable a unified and robust output for multi-distribution prediction.
By stacking multiple layers, MDL enables feature, scenario, and task information to fully interact in a bottom-up and layer-wise manner.
This deep integration allows the model's massive parameter space to be dynamically steered by distribution-specific prompts, resulting in an effective modeling of complex multi-distribution data.

To evaluate the proposed approach \modelname, we conducted extensive experiments on real-world industrial dataset.
The experimental results demonstrate that our approach can achieve much better performance than competitors for multi-scenario and multi-task recommendations in various settings.
Furthermore, we conducted an online A/B test on Douyin Search for one month, achieving a significant improvement of +0.0626\% in LT30 and -0.3267\% in change query rate.
And the proposed approach MDL has been fully deployed on Douyin Search.

\section{PRELIMINARIES}
\label{sec-pre}

\begin{figure*}[ht]
    \centering
    \includegraphics[width=1.0\textwidth]{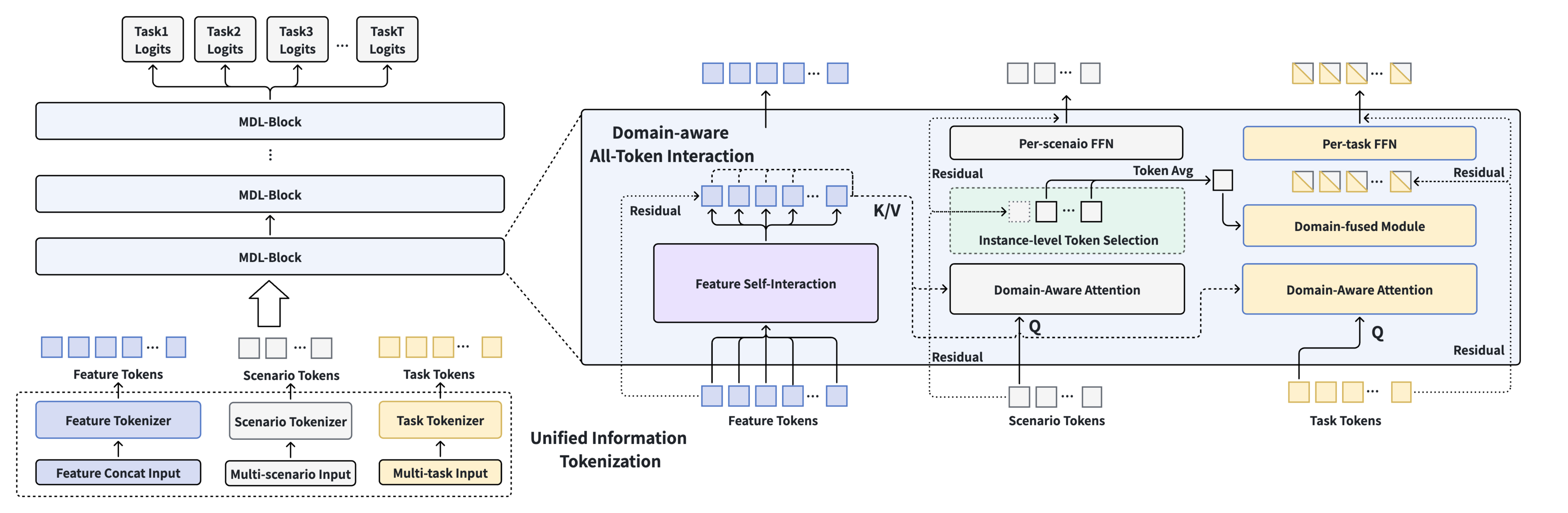}
  \caption{The overall framework of our approach MDL.}
    \label{fig:model-overall}
\end{figure*}

We first formulate the multi-scenario and multi-task recommendation problem.
Let $\mathcal{X}$ and $\mathcal{Y}$ denote the feature space and the label space, respectively.
The feature space usually consists of user features, sequence features, item features and cross features.
And the label space represents different prediction tasks $\mathcal{Y} = \{\mathcal{Y}^{1},\cdots,\mathcal{Y}^{N}\}$ (\eg click or like), where $N$ is the number of tasks.
Consider $K$ different scenarios $\{\mathcal{S}^1, \cdots, \mathcal{S}^K\}$, where the instance set of scenario $k$ is represented as $\mathcal{S}^k=\{(x_i^k,y_i^{1,k},y_i^{2,k},\cdots,y_i^{N,k})\}_{i=1}^{|\mathcal{S}^k|}$. 
Here $x_i^k \in \mathcal{X}$ is the feature representation of the $i$-th instance, and $y_i^{n,k}$ ($n \in N$) is the label of the $n$-th task.
Multi-scenario and multi-task recommendation methods aim to leverage all scenarios' samples to train a model $\hat{y}_{i}^{n,k}=p_{\Theta}(x_{i}^{k})$ (parameterized by $\Theta$) that can accurately predict each user behavior label $y_i^{n, k}$.

Different tasks employ distinct optimization functions to optimize model parameters $\Theta$, and the standard loss function of multi-scenario and multi-task recommendation can be formulated as follows:

\begin{equation}
\label{eq:main_loss}
    \mathcal{L}_{rec} = \sum_{k=1}^{K}\sum_{n=1}^{N}\sum_{i=1}^{|\mathcal{S}^{k}|}\ell(y_{i}^{k}, \hat{y}_{i}^{n,k}),
\end{equation}
where $\ell(\cdot)$ is the loss function, usually is the cross-entropy loss.

\section{Methodology}
\label{sec-model}
In this section, we present the \textbf{M}ulti-\textbf{D}istribution \textbf{L}earning (\textbf{MDL}) framework, a unified architecture designed for multi-scenario learning and multi-task learning in recommendations.
The core idea of our approach is "Tokenize-and-Interact".
As shown in Figure~\ref{fig:model-overall}, our approach mainly consists of two parts: \emph{Unified Information Tokenization} module and \emph{Domain-aware All-Token Interaction} module.
The first module is to transform heterogeneous inputs into a set of discrete latent tokens, \ie feature tokens, scenario tokens, and task tokens.
The second module describes how these tokens interact for achieving unified and effective modeling across multiple scenarios and tasks.
Next, we describe each part in detail.

\subsection{Unified Information Tokenization}
Firstly, we introduce how to represent feature, task and scenario information in a tokenized form uniformly.

\subsubsection{Feature Tokenization}
The input feature is the fundamental of a reommendation model.
In real-world industrial recommendation systems, input features typically include user features $x_u$ (\eg user id, user profile), item features $x_i$ (\eg video id, author id), sequence features $x_{seq}$ (\eg user history click behaviors), and cross features $x_{cross}$. 
Most user features, item features and cross features are first transformed into high-dimensional embeddings $\bm{e}$ through the embedding layers as follows:
\begin{equation}
\bm{e}_{u}, \bm{e}_{i}, \bm{e}_{cross} = \text{EmbLayer}(x_u,x_i,x_{cross}).
\end{equation}
As for sequence features $x_{seq}$, they are usually processed through the designed sequence modules~\cite{DIN,Longer,SIM} to extract user interest and convert them into embeddings $\bm{e}_{seq}$. 
Ultimately, all features are transformed into feature embeddings $\bm{e}_{input} = [\bm{e}_u; \bm{e}_i; \bm{e}_{cross}; \bm{e}_{seq}]$ with varying dimensions.

To achieve complex feature interaction(\eg Self-Attention~\cite{transformer}, MLPMixer~\cite{MLPMixer}), existing large-scale recommendation methods typically transform features into the fixed-number and fixed-dimensional hidden representations $\bm{T}_{f} \in \mathbb{R}^{N_{f} \times d_{f}}$, referred to as feature tokens~\cite{RankMixer,OneTrans,KGBRank}. 
These feature tokens ensure efficient parallel computation and effective feature interaction in later stages.
Following previous work~\cite{RankMixer}, we adopt a semantic-based tokenization approach for generating feature tokens in \modelname.
Specifically, we manually group features into serval semantically coherent clusters with domain knowledge.
These grouped features are sequentially concatenated as the input embeddings $\bm{e}_{input} = [\bm{e}_1; \bm{e}_2; \cdots; \bm{e}_{N_{f}}]$, where $N_{f}$ is the number of feature tokens.
For each grouped feature embedding $\bm{e}_j$, it captures a set of feature embeddings that represent a similar semantic aspect.
Then a projection layer convert the grouped feature embedding $\bm{e}_j$ into a fixed-dimension hidden representation as follows:
\begin{equation}
    \bm{t}_{j} = \text{Proj}(\bm{e}_{j}),
\end{equation}
where $\bm{t}_{j} \in \mathbb{R}^{d_{f}}$, $d_{f}$ is the dimension of the feature token.
Finally, we obtain the feature tokens $\bm{T}_{f} = [\bm{t}_{1};\bm{t}_2;\cdots;\bm{t}_{N_{f}}]$.

\subsubsection{Scenario and Task Tokenization}
Inspired by large language models (LLMs), to further enhance the interaction between feature, scenario,  and task information, we propose tokenizing scenario information and task information to align them with the feature tokens in the same semantic space.
Compared with previous multi-scenario and multi-task recommendation methods, a unified tokenization framework enables multi-scenario and multi-task signals to interact with feature information from the bottom layer and progressively across layers, rather than affecting solely on partial intermediate results or shallow output results.
Similar to how LLMs are applied to specific downstream tasks, these scenario and task tokens act as prompt tokens to activate the model's capabilities for multi-scenario and multi-task recommendation.

Next, we will introduce the tokenization approach for scenarios and tasks.
We begin by taking scenario tokens as an example.
Similar to feature tokens $\bm{T}_{f}$, scenario tokens $\bm{T}_{t} \in \mathbb{R}^{(N_{s} + 1) \times d_{s}}$ are also a fixed number of hidden representations, which are derived from specific feature embeddings through a transformation process.
Specifically, the input feature embeddings for scenario tokens consist of two parts: (1) some important features' extra embeddings $\bm{\hat{e}_{imp}}$ (\eg user id, video id) where is different from the input embeddings of feature tokens $\bm{e}$, but same raw features $x$, and (2) scenario-related prior feature embeddings $\bm{\hat{e}_{spec}}$ (\eg scenario-specific user behavior sequence).
The two components ensure the richness of the initial input information and the differentiation among the different scenario tokens.
Then, the nolinear feed-forward network is used to convert the input embeddings into task tokens as follows:
\begin{equation}
    \bm{t}_{s} = \text{Relu}(\text{FFN}(\bm{\hat{e}_{imp}} \oplus \bm{\hat{e}_{spec}})).
\end{equation}
For each scenario token $\bm{t}_{s}$, the used FFN parameters are different, \ie per-token feed-forward network (Pertoken FFN)~\cite{RankMixer}.
Finally, we obtain the scenario tokens $\bm{T}_{s} = [\bm{t}_{s,1};\bm{t}_{s,2};\cdots;\bm{t}_{s,(N_{s} + 1)}] \in \mathbb{R}^{(N_{s} + 1) \times d_{s}}$, where $d_{s}$ is the dimension of the scenario token, $N_{s}$ is the number of scenario tokens, which equals to the number of recommendation scenarios (\eg homepage feeds or banner).
In addition to the $N_{s}$ scenario tokens, we also construct a global scenario token $\bm{t}_{s,global} \in T_{s}$ that is shared across all scenarios.
The global scenario token is used to learn common knowledge among scenarios.

Similarly, we can obtain task tokens $\bm{T}_{t} \in \mathbb{R}^{N_{t} \times d_{t}}$ by inputting other important feature embeddings and task-related feature embeddings, and using Pertoken FFN transformation.
$N_{t}$ is the number of task tokens, which is equal to the number of prediction tasks of the recommendation model.

\subsection{Domain-aware ALL-Token Interaction}
After obtaining the initial feature tokens, scenario tokens and task tokens, we introduce how to optimize token information by controlling the interaction and information transmission among different tokens, further achieving effective and unified multi-scenario and multi-task recommendation modeling.
The interactions among these tokens can be categorized into three main types: \emph{feature token self-interaction}, \emph{feature-scenario/task token interaction} and \emph{scenario-task token interaction}.
Next, we will introduce in detail.

\subsubsection{Feature Token Self-Interaction}
The interaction between feature tokens serves as a fundamental module in large-scale industrial recommendation models, which has been widely studied in recent works~\cite{OneTrans, RankMixer, hyformer}.
It aims to enhance the model's expressive ability by stacking efficient feature interaction modules to scale model parameters.
For instance, structures like Self-Attention~\cite{transformer} and MLP-Mixer~\cite{MLPMixer}, along with their variants, have been widely applied in the design of feature interaction modules.
The feature interaction layer can be formulated as follows:
\begin{equation}
    \bm{T}_{f}^{(l+1)} = \text{Sef-Interaction}(\bm{T}_{f}^{(l)}),
\end{equation}
where $l$ is the number of layers, $\bm{T}_{f}^{(l)}, \bm{T}_{f}^{(l+1)} \in \mathbb{R}^{N_{t} \times d_{t}}$ represent the feature tokens before and after interaction, respectively.
Specifically, in this work, we follow the feature interaction module design from the previous approach RankMixer~\cite{RankMixer}.
It mainly consists of two parts: TokenMixing module and Pertoken FFN.
So, the feature token self-interaction in \modelname is as follows:
\begin{equation}
\label{eq:feature-interaction}
    \bm{T}_{f}^{(l+1)} = \text{PertokenFFN}(\text{LN}(\text{TokenMixing}(\bm{T}_{f}^{(l)}) + \bm{T}_{f}^{(l)})).
\end{equation}
It is worth noting that this part can be replaced by other feature interaction methods.

For multi-scenario and multi-task learning in recommender system, the interaction among feature tokens provides the foundational modeling capability.
Next, we introduce how the defined scenario tokens and task tokens can be utilized to better activate these capabilities.

\subsubsection{Feature-Scenario/Task Token Interaction}
The interaction between scenario/task tokens and feature tokens primarily aims to enable different scenario tokens or task tokens to utilize distinct feature information, thereby achieving differentiated modeling in multi-scenario and multi-task settings.
Since task and scenario information are tokenized from the bottom layer, they can fully interact with feature information layer by layer and from the bottom up.
This approach alleviates the challenge of insufficient utilization of complex feature interaction modules, which was previously limited to the certain intermediate or output layers.

Specifically, we define a type of interaction called \emph{Domain-aware Attention} between scenario/task tokens and feature tokens.
Without loss of generality, we take the interaction between task tokens $\bm{T}_{t}$ and feature tokens $\bm{T}_{f}$ as an example to illustrate this module.
The \emph{Domain-aware Attention} can be viewed as a variant of cross-attention.
Given the task token $\bm{T}_{t}$, we use it as query, while treating the feature token $\bm{T}_{f}$ as key and value.
Then they are interacted through cross multi-head attention as follows:
\begin{equation}
\label{eq-daa-task}
    \bm{\hat{T}}_t^{(l+1)} = \text{softmax}(\frac{(\bm{W}_Q\bm{T}_t^{(l)})(\bm{W}_K\bm{T}_{f}^{(l)})}{\sqrt{d}})\bm{W}_V\bm{T}_f^{(l)},
\end{equation}
where $l$ is the number of layers, $\bm{W}_Q, \bm{W}_K, \bm{W}_V$ are the QKV projection matrices in the form of a PerToken FFN.
Similarly, for the scenario token $\bm{T}_{s}$, we use it as the query and perform the same multi-head cross-attention interaction with the feature token $\bm{T}_{f}$:
\begin{equation}
\label{eq-daa-scenario}
    \bm{\hat{T}}_s^{(l+1)} = \text{softmax}(\frac{(\bm{W}_Q\bm{T}_s^{(l)})(\bm{W}_K\bm{T}_{f}^{(l)})}{\sqrt{d}})\bm{W}_V\bm{T}_f^{(l)}.
\end{equation}
Through the cross-attention interaction between tokens, specific task tokens and scenario tokens can adaptively and selectively aggregate the rich information from feature tokens based on their own data characteristics, enabling differentiated modeling for different tasks and scenarios. 
Furthermore, since tokenization is performed from the bottom input layer, this interaction can be stacked across multiple layers with feature token self-interaction layer.
The layer-wise cross-attention interaction ensures effective utilization of information within complex structures.

\subsubsection{Scenario-Task Token Interaction}
\label{subsubsec-interaction3}
The interaction between task tokens and scenario tokens primarily focuses on the fusion of their information, enabling differentiated model predictions for specific task within specific scenario in the context of tokenization.
This interaction is mainly controlled by the \emph{Domain-fused Module}.
The core idea is that for each instance, we aim to aggregate only the corresponding scenario token information $\bm{t}_{s}$ for each task token $\bm{t}_{t}$.
This ensures scenario-specific prediction for different recommendation tasks.

Specifically, the \emph{Domain-fused Module} consists of two steps: (1) instance-level scenario token selection, and (2) hybrid information fusion between task and scenario tokens.
Firstly, for each instance in the dataset, we select the corresponding scenario tokens $\{{\bm{t}_{s,i}, \bm{t}_{s,j}}\}$ from the full scenario token set $\bm{T}_s = \{\bm{t}_{s,1}, \bm{t}_{s,2}, \cdots, \bm{t}_{s,{n_s}}, \bm{t}_{s, global}\}$ based on the scenario information associated with each instance, along with the global scenario token $\bm{t}_{s, global}$.
For example, suppose a model is designed to serve $N_{s}$ scenarios. 
For a given instance, it must belong to one or more of these scenarios (in cases where scenario overlap exists). 
Based on this information, the corresponding scenario tokens are selected. 
After selecting the scenario tokens $\{{\bm{t}_{s,i}, \bm{t}_{s,j}}, \bm{t}_{s,global}\}$, we aggregate their information using a mean pooling approach:
\begin{equation}
\label{eq-df1}
    \bm{t}_{s, avg} = \text{MeanPooling}(\{\bm{t}_{s,i}, \bm{t}_{s,j},\bm{t}_{s,global}\}),
\end{equation}
where $\bm{t}_{s, avg}$ represents the aggregated scenario information corresponding to the instance.
Next, we fuse the aggregated scenario token information $\bm{t}_{s, avg}$ into the task token $\bm{T}_{t}$.
For each Task token $\bm{t}_{t,k} \in \bm{T}_{t}$, we directly use a sum pooling approach to integrate the two pieces of information:
\begin{equation}
\label{eq-df2}
    \bm{T}_{t} = \bm{T}_{t} + \bm{t}_{s,avg}.
\end{equation}
In our practice, the simple pooling approach is not only efficient but also sufficiently effective.
Through this tokenized information fusion, the prediction between scenarios and tasks is decoupled, enabling flexible combinations of any scenario and task modeling.

\subsection{MDL Block}
After introducing the main types of interactions in MDL, we now provide a more detailed description of the interaction process within a complete MDL block by using these interactions.

\paratitle{Forward Propagation of Feature Tokens.}
The information propagation of feature tokens primarily relies on their self-interaction.
For each layer of the MDL Block, the propagation of feature tokens is equal to Equation~\ref{eq:feature-interaction}.

\paratitle{Forward Propagation of Scenario Tokens.}
The information propagation of scenario tokens primarily involves domain-aware attention between scenario tokens and feature tokens (Eq~\ref{eq-daa-scenario}), per-scenario FFN, and residual connection.
It can be formulated as follows:
\begin{equation}
    \bm{\hat{T}}_{s}^{(l + 1)} = \text{DomainAwareAttn}(\bm{T}_{s}^{(l)}, \bm{T}_f^{(l + 1)}) + \bm{T}_{s}^{(l)},
\end{equation}
\begin{equation}
    \bm{T}_{s}^{(l + 1)} = \text{PertokenFFN}(\bm{\hat{T}}_{s}^{(l + 1)}) + \bm{\hat{T}}_{s}^{(l + 1)}.
\end{equation}
The per-scenario FFN after the domain-aware attention provides additional nonlinear transformations for each scenario token, enhancing the expressive capability of each scenario token.
And the residual connection ensures the training stability of the scenario tokens and the continuity of information across layers.

\paratitle{Forward Propagation of Task Tokens.}
The information propagation of task tokens primarily consists of domain-aware attention between task tokens and feature tokens (Eq~\ref{eq-daa-task}), domain-fused module (Eq~\ref{eq-df1}, Eq~\ref{eq-df2}), per-task FFN and residual connection.
It can be formulated as follows:
\begin{equation}
    \bm{\hat{T}}_{t}^{(l + 1)} = \text{DomainAwareAttn}(\bm{T}_{t}^{(l)}, \bm{T}_f^{(l + 1)}) + \bm{T}_{t}^{(l)},
\end{equation}
\begin{equation}
    \bm{\tilde{T}}_t^{(l+1)} = \text{DomainFusedModule}(\bm{\hat{T}}_{t}^{(l + 1)}, \bm{\hat{T}}_{s}^{(l + 1)}),
\end{equation}
\begin{equation}
    \bm{T}_t^{(l+1)} = \text{PertokenFFN}(\bm{\tilde{T}}_t^{(l+1)}) + \bm{\tilde{T}}_t^{(l+1)}.
\end{equation}
The pertoken FFN and residual connection are also used in the information propagation of task tokens.

Finally, by stacking $L$ layers of MDL Blocks, we obtain the final output task tokens $\bm{T}_t^{(L)}$.
We attach the logits layer to the final layer of task tokens to serve as the output for the model's different prediction task as follows:
\begin{equation}
    \hat{y}_{n} = \text{LogitsLayer}(\bm{t}_{t,n}^{(L)}),
\end{equation}
where $n \in N_{t}$ is the index of the prediction task.
Since the final output task tokens have already aggregated the corresponding scenario token information for each instance through the domain-fused module, the gradients of instances from a specific scenario will only update the corresponding scenario tokens. 
In this way, scenario tokens are influenced solely by their own scenario information, allowing them to aggregate information from feature tokens and effectively model the differences between scenarios.
And because the tokenized information aggregation is performed bottom-up, the number of final prediction outputs is independent of the scenarios, yet it enables differentiated predictions across scenarios.

\section{EXPERIMENTS}
\label{sec-exp}

\subsection{Experimental Setup}
We first describe the experimental setup, including datasets, evaluation settings, comparison methods and implementation details.

\subsubsection{Datasets.}
To evaluate the effectiveness of the proposed approach, we use a large-scale production dataset from \emph{Douyin search system}.
The dataset is collected from the real online logs including 3 main search sceanrios (\ie \emph{single-column search}, \emph{double-column search}, and \emph{inner search}) and more than 20 prediction tasks.
Specifically, we collect 2-month consecutive user interaction
logs involving billions of users and hundreds of millions of documents.
Each instance incorporates over 500 features including user features, item features, sequence features and cross features.
Besides, we keep 1\% of the dataset as the evaluation data, which are excluded from the training process.
All datasets used for training and evaluation are strictly anonymized to protect user privacy.

\begin{table*}[h]
    \centering
    \caption{Performance comparison of different methods on the production dataset. The best performance and the runner-up performance are denoted in bold and underlined fonts respectively. "Improv." indicates the relative improvement ratios of the proposed approach over the best performance baselines.}
	\label{tab:results-all}
	\begin{tabular}{l c c c c c c c c c c c}
		\toprule
		\multirow{2}{*}{Method}&\multicolumn{3}{c}{Single-column Search}&&\multicolumn{3}{c}{Double-column Search}&&\multicolumn{3}{c}{Inner Search}\\
		\cmidrule{2-4}
		\cmidrule{6-8}
		\cmidrule{10-12}
		& $\text{QAUC}_{\text{Click}}$ & $\text{QAUC}_{\text{Like}}$ & $\text{QAUC}_{\text{Fav}}$ && $\text{QAUC}_{\text{Click}}$ & $\text{QAUC}_{\text{Like}}$ & $\text{QAUC}_{\text{Fav}}$ && $\text{QAUC}_{\text{Click}}$ & $\text{QAUC}_{\text{Like}}$ & $\text{QAUC}_{\text{Fav}}$\\
        \midrule
        RankMixer & 0.6389 & 0.6610 & 0.6624 && 0.6911 & 0.6621 & 0.6787 && 0.6393 & 0.6786 & 0.6621 \\
        \midrule
        SharedBottom & 0.6391 & 0.6611 & 0.6614 && 0.6929 & 0.6629 & 0.6789 && 0.6395 & 0.6794 & 0.6609\\
        MMoE & 0.6394 & 0.6614 & 0.6623 && \underline{0.6931} & \underline{0.6629} & 0.6788 && 0.6397 & 0.6793 & 0.6629 \\
        STAR & 0.6389 & 0.6611 & 0.6600 && 0.6930 & 0.6624 & 0.6784 && 0.6393 & 0.6796 & 0.6631 \\
        HMoE & \underline{0.6397} & \underline{0.6617} & 0.6623 && 0.6925 & 0.6625 & \underline{0.6790}  && 0.6401 & 0.6790 & \underline{0.6634} \\
        PEPNet & 0.6396 & 0.6609 & \underline{0.6628} && 0.6928 & 0.6625 & 0.6787  && \underline{0.6403} & \underline{0.6797} & 0.6627 \\
		\midrule
		MDL (Our) & \textbf{0.6417} & \textbf{0.6632} & \textbf{0.6656} && \textbf{0.6950} & \textbf{0.6671} & \textbf{0.6842} && \textbf{0.6419} & \textbf{0.6820} & \textbf{0.6681} \\
        Improv. & \emph{+0.31\%} & \emph{+0.23\%} & \emph{+0.42\%} && \emph{+0.27\%} & \emph{+0.63\%} & \emph{+0.77\%} && \emph{+0.25\%} & \emph{+0.34\%} & \emph{+0.71\%} \\
		\bottomrule
	\end{tabular}
\end{table*}

\subsubsection{Evaluation Settings.}
\label{subsec-evluation}
For offline evaluation, we adopt the QAUC (Query-level AUC) as the primary performance metrics, which reflects the model's ranking ability on candidates, and is widely used in personalized search system~\cite{cotrain,hyformer}.
Specifically, QAUC calculates the AUC for samples within each query and then averages the results across all queries.
It is calculated as follows:
\begin{equation}
\label{eq:qauc}
    \text{QAUC} = \frac{\sum_{i=1}^{{N}} \text{AUC}_{i}}{{N}},
\end{equation}
where $N$ is the number of distinct UID-query pairs.
To evaluate the performance across multiple scenarios and tasks, we selected three primary prediction tasks as representatives: \emph{click}, \emph{like}, and \emph{favorite}, and evaluate the performance of these prediction objectives across three different scenarios on the used production dataset.

For online evaluation, we adopt \emph{30-day user lifetime} (\ie LT30) and \emph{change query rate}.
LT30 refers to the average number of active days per user in the latest 30days, which is a important metric for DAU growth.
The \emph{change query rate} measures the probability of users manually refining a search query into a more specific one (\eg changing from "tank" to "tank300"). It is calculated as follows:
\begin{equation}
\label{eq:changeq}
    Ratio_{change} = \frac{{\tilde{N}}_\text{reform}}{{N}_\text{total}},
\end{equation}
where $\tilde{{N}}_\text{reform}$ is the number of distinct UID-query pairs with query reformulation.
This metric measures user satisfaction with the search experience, which is also widely used in personalized search system~\cite{cotrain,hyformer}.
A lower value indicates a better user experience and demonstrates a better performance of the model's predictions.

\subsubsection{Comparison Methods.}
In order to verify the effectiveness of the proposed approach, we mainly consider the following methods:

$\bullet$ \textbf{RankMixer}~\cite{RankMixer}: It adopts the token mixing and pertoken FFN strategies to capture heterogeneous feature interaction, which achieve remarkable scaling law in recommendation methods.
We use this method as a representative backbone baseline without incorporating additional multi-task or multi-scenario designs.

$\bullet$ \textbf{SharedBottom}~\cite{sharedbottom}: It adopts a classic \emph{shared-specific} framework that shares the parameters of the bottom layer and designs scenario-specific and task-specific parameter tower for each scenario and recommendation task.

$\bullet$ \textbf{MMoE}~\cite{MMoE}: It designs the shared multi-gate Mixture-of-Experts (MoE) structure to model task and scenario relationships and capture common representations from multiple scenarios and tasks.

$\bullet$ \textbf{STAR}~\cite{STAR}: It shares a centered network for all the domain and generates domain-specific network for each scenario and task.

$\bullet$ \textbf{HMoE}~\cite{HMoE}: It proposes the hybrid of implicit and explicit Mixture-of-Experts approach which learns the scenario and task relationships from feature space implicitly and label space explicitly.

$\bullet$ \textbf{PEPNet}~\cite{PEPNet}: It proposes a parameter and embedding personalized network to learn the heterogeneous relationships between multiple scenarios and multiple tasks.

Among them, except for RankMixer, all others are specifically designed for multi-scenario and multi-task recommendation models.
Therefore, in practical implementation, we integrate these methods into the RankMixer backbone, and ensure the total number of model parameters remains consistent for performance comparison.

\subsubsection{Implementation Details.}
All experiments are conducted on hundreds of GPUs in a hybrid distributed training framework that the sparse part is updated asynchronously, while the dense part is updated synchronously.
The optimizer hyperparameters are kept consistent acrossall models.
For the dense part, we used the RMSProp optimizer, while the sparse part used the Adagrad optimizer.
The batch size is set to 2048.
By adjusting the number of layers and the hidden layer dimensions of different models, we ensure that the total parameter size of all models is controlled at around 0.5B.

\subsection{Performance Comparison}
We compare the proposed approach with the aforementioned baselines on the adopted datasets, and the results are summarized in Table~\ref{tab:results-all}.
From it, we have the following observations:

Compared with Rankmixer which is a representative backbone baseline without any multi-task and multi-scenario designs, almost all the multi-scenario and multi-task baselines (SharedBottom, MMoE, STAR, HMoE and PEPNet) achieve better performance.
Because the designed shared-specific structure in these methods can better capture commonalities and characteristics between different scenarios and tasks.
Further, by comparing the proposed approach MDL with all the baselines, it is clear that MDL consistently performs better than them by a large margin on both scenarios and tasks.
It is because the proposed approach leverages tokenization to enable more comprehensive interactions among scenario, task, and feature information.

Furthermore, the proposed approach MDL demonstrates superior performance in scenarios with sparser data and tasks with lower positive sample rates.
Compared with the improvement in the single-column search, the proposed method shows even better results in the double-column search and inner search scenarios, which have sparser data.
Additionally, compared to the click task, the proposed method achieves better results in the like and favorite tasks, which have lower positive sample rates than click task.
We believe that tokenized information representation and token-level interactions can further mitigate the seesaw effect in multi-distribution learning.
During the training process, this approach allows the model to better maintain distribution modeling for relatively disadvantaged scenarios and tasks.

\begin{table}[htbp]
    \centering
    \caption{Ablation study of the proposed approach on the three scenarios ($\triangle\text{QAUC}_{\text{Click}}$).}
    \label{tab:ablation}
        \begin{tabular}{l c c c}
            \toprule
            Variants & Single & Double & Inner\\
            \midrule
            \textbf{Ablation of Task Token} \\
            \midrule
            $w/o$ task token & -0.12\% & -0.11\% & -0.09\% \\
            $w/o$ task-feature interaction & -0.04\% & -0.05\% & 0.03\% \\
            \midrule
            \textbf{Ablation of Scenario Token} \\
            \midrule
            $w/o$ scenario token & -0.17\% & -0.16\% & -0.15\% \\
            $w/o$ global scenario token & -0.04\% & -0.06\% & -0.05\% \\
            $w/o$ scenario-feature intraction & -0.05\% & -0.04\% & -0.05\% \\
            \bottomrule
        \end{tabular}
\end{table}

\subsection{Ablation Study}
In this part, we continue to analyze how each of the proposed techniques or components affects the final performance.
We prepare five variants of the proposed approach MDL for comparisons, including:

$\bullet$ $w/o$ task token: without all task tokens from the input, and leverage task-tower for multi-task learning.

$\bullet$ $w/o$ task-feature interaction: replace domain-aware attention on task tokens with interaction in Rankmixer.

$\bullet$ $w/o$ scenario token: without all scenario tokens from the input, and leverage scenario-tower for multi-scenario learning.

$\bullet$ $w/o$ global scenario token: without the global scenario token from the input.

$\bullet$ $w/o$ scenario-feature interaction: replace domain-aware attention on scenario tokens with interaction in Rankmixer.

The experimental results of the MDL variants are reported in Table~\ref{tab:ablation}.
From it, we can observe that removing each of the components leads to the performance decrease.
The variant $w/o$ task token and $w/o$ scenario token indicate the importance of tokenization of scenario information and task information.
The variant $w/o$ global scenario token shows the importance of the global token.
The variants $w/o$ task-feautre interaction and $w/o$ scenario-feature interaction indicate the effectiveness of the elaborated interaction mechanisms among these tokens.

\subsection{Further Analysis}
In this section, we further perform a series of detailed analyses on
the proposed MDL to confirm its effectiveness.

\begin{figure}[t]
    \centering
    \subfigure[QAUC scaling with model params.]{
        \centering
        \includegraphics[width=0.225\textwidth]{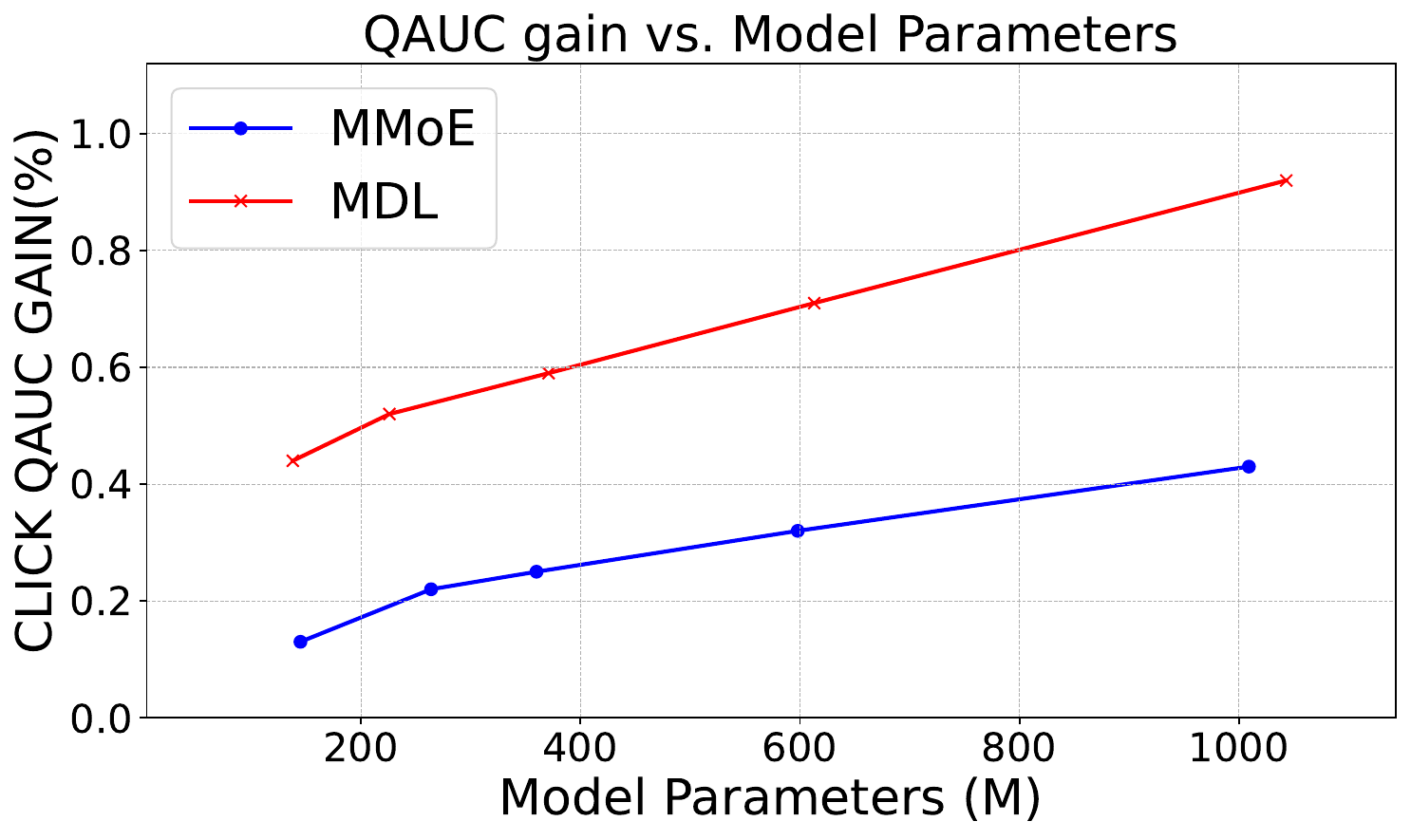}
    }
    \subfigure[QAUC scaling with model FLOPs.]{
        \centering
        \includegraphics[width=0.225\textwidth]{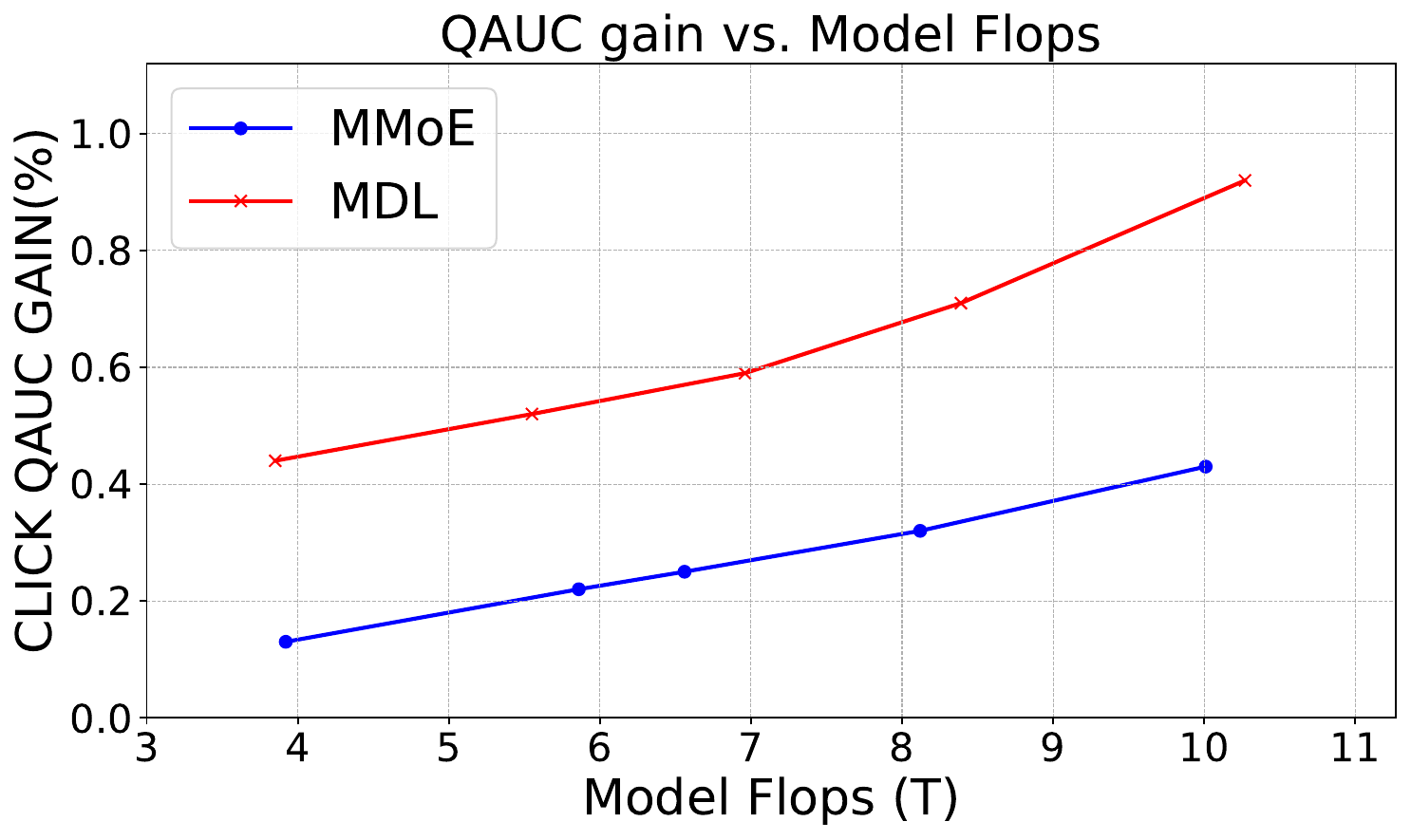}
    }
\caption{Scaling laws between click QAUC gain on single-column search scenario and model parameters/FLOPs of different models.}
\label{fig:exp-sl}
\end{figure}

\subsubsection{Performance Comparison \wrt Scaling Laws}
In this part, we examine how the proposed approach performs \wrt the model parameters and model FLOPs.
For this purpose, we adjust the model's hidden representations $d$ and the number of layers $L$ to control the model's parameter size and FLOPs.
Then we compare the performance of MDL and the representive baseline MMoE on these settings, and report the results in Figure~\ref{fig:exp-sl}.
From this, we can find that the performance of MDL is consistently better than MMoE.
Meanwhile, as the model params and FLOPs increase, the performance gain brought by MDL increases.
Compared with previous multi-scenario and multi-task recommendation methods, the proposed approach can better leverage the model parameters for multi-distribution learning through the token-level interaction layer-by-layer, fully harvesting the benefits of the scaling law.

\begin{figure}[t]
    \centering
    \subfigure[Click task on layer1.]{
        \centering
        \includegraphics[width=0.225\textwidth]{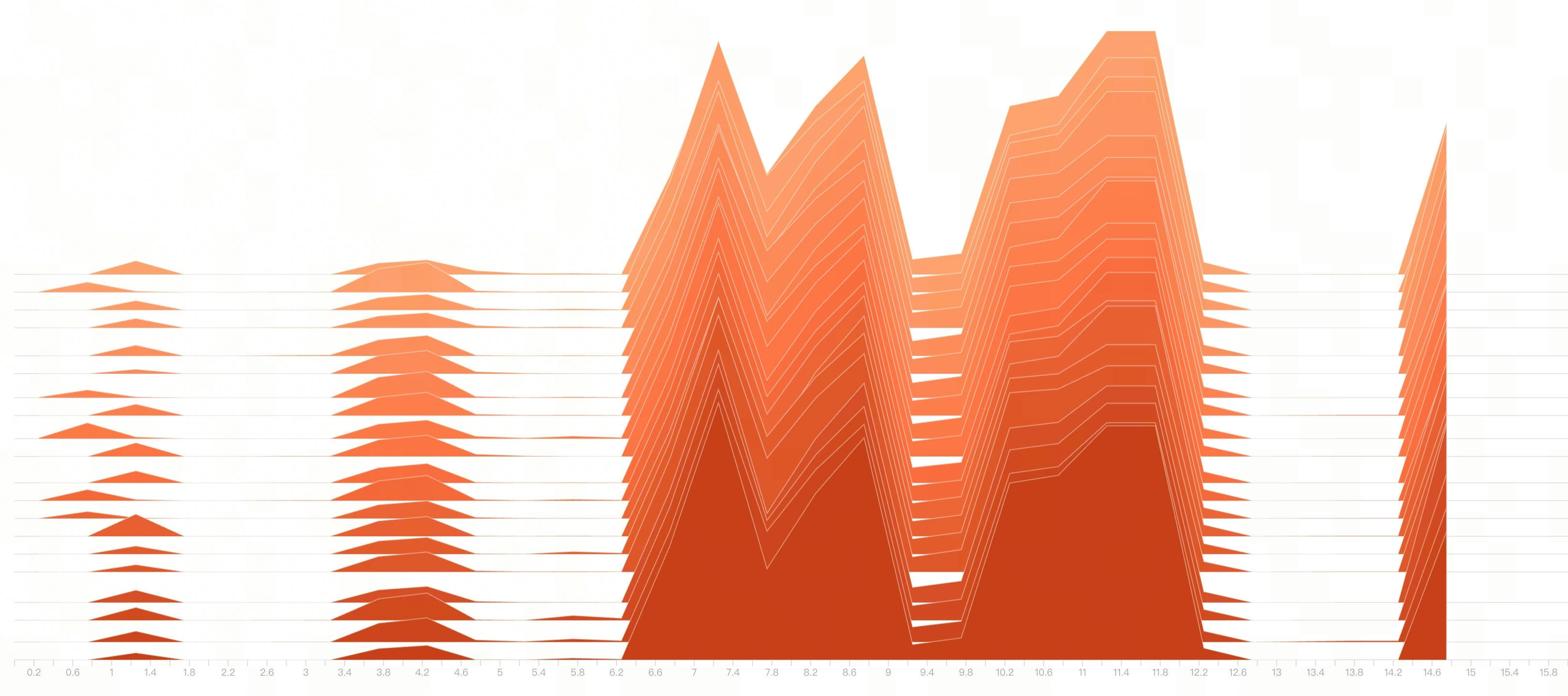}
    }
    \subfigure[Click task on layer2.]{
        \centering
        \includegraphics[width=0.225\textwidth]{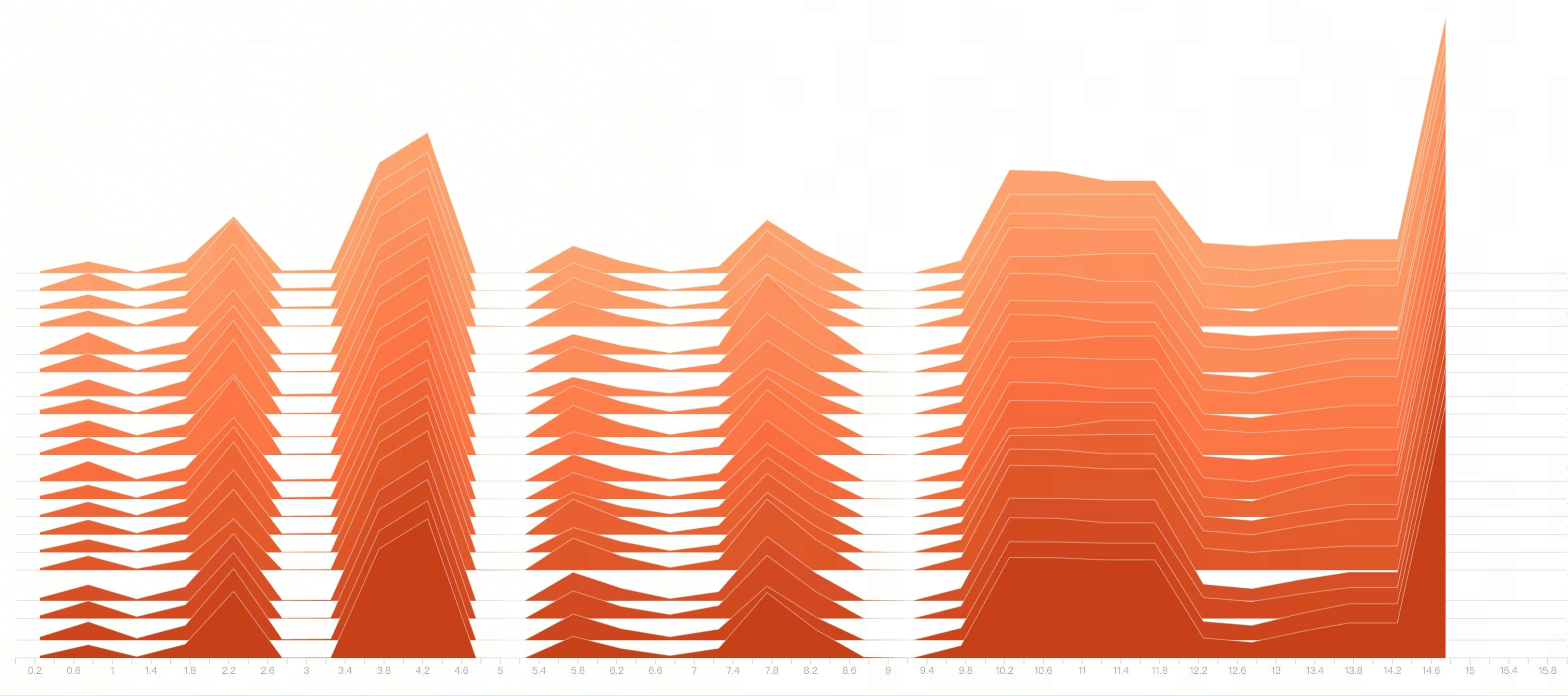}
    }
    \subfigure[Like task on layer1.]{
        \centering
        \includegraphics[width=0.225\textwidth]{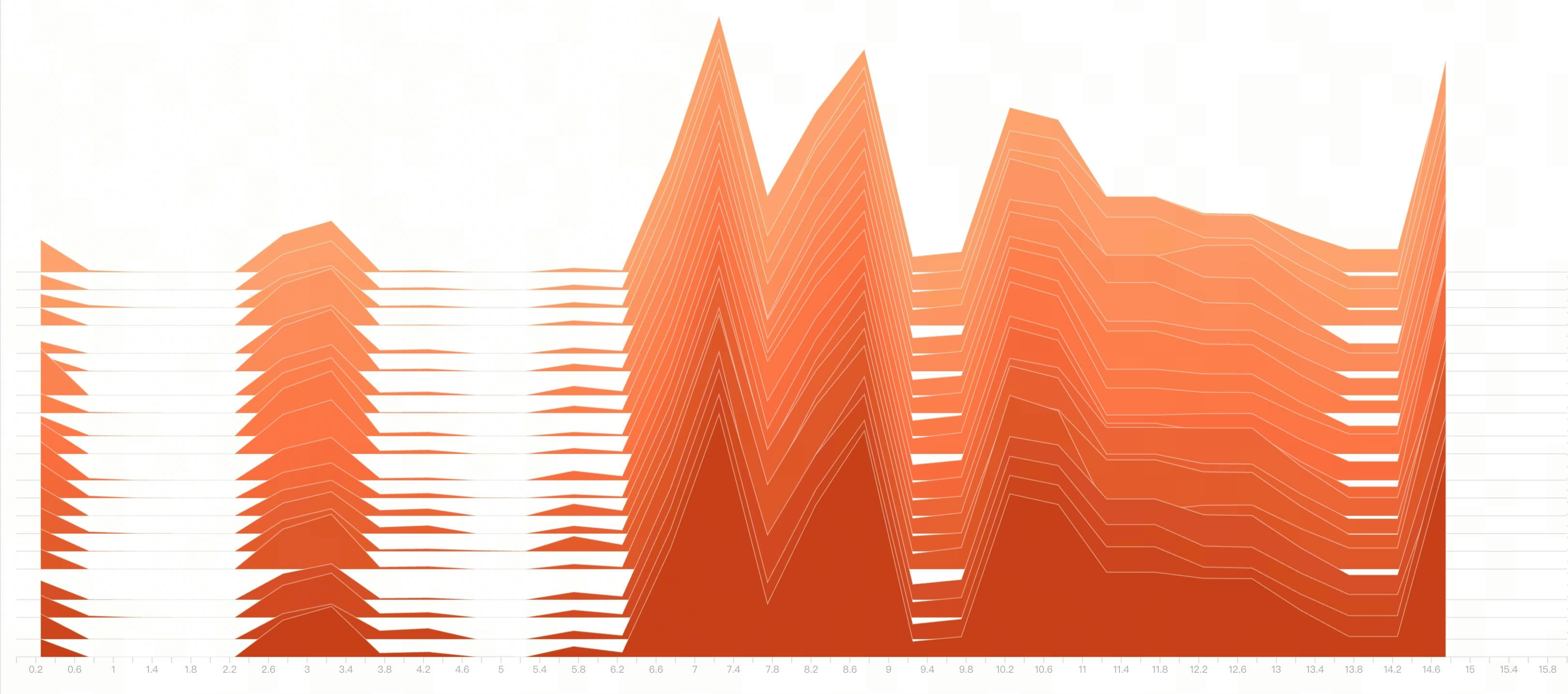}
    }
    \subfigure[Like task on layer2.]{
        \centering
        \includegraphics[width=0.225\textwidth]{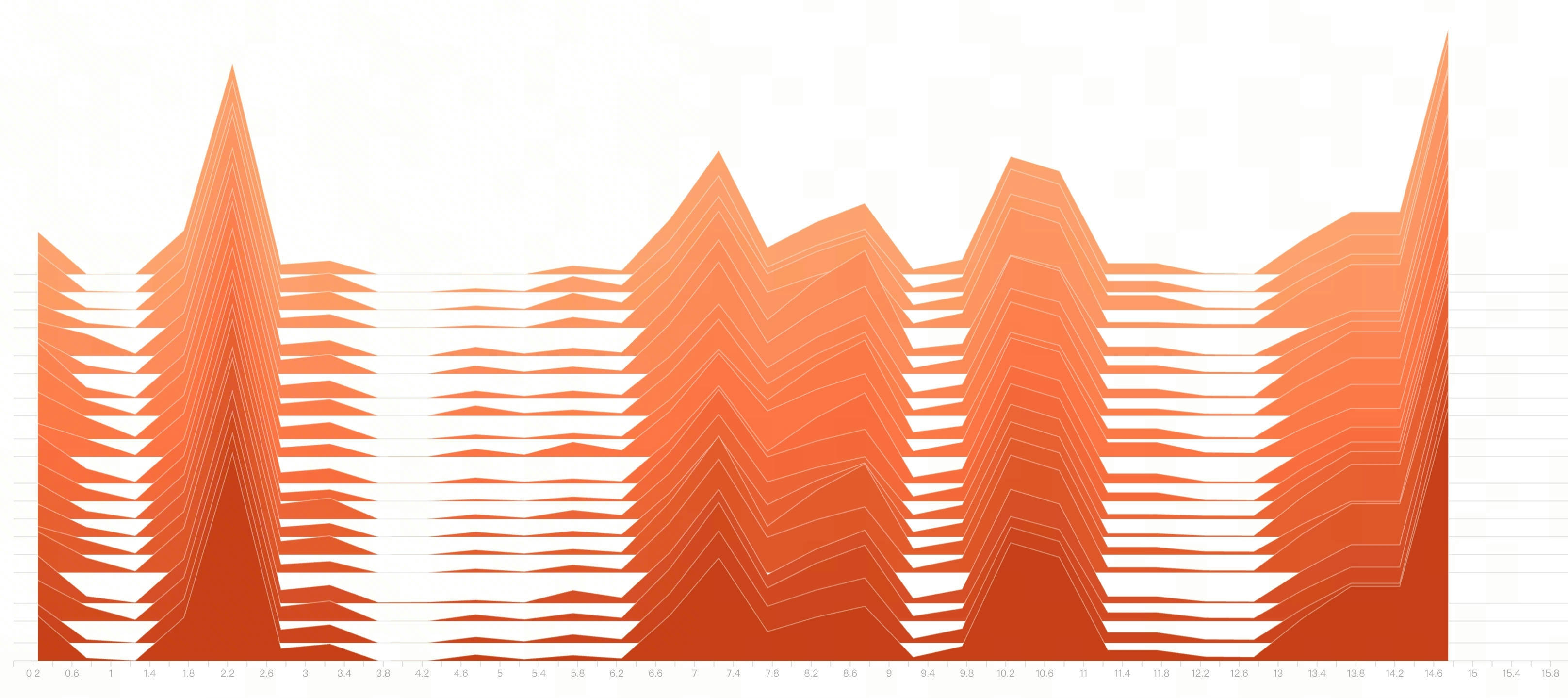}
    }
\caption{Attention distribution between task tokens and feature tokens on different layers. The x-axis represents the index of feature tokens and y-axis represents the average attention weights belong to the corresponding feature token.}
\label{fig:exp-vis1}
\end{figure}

\begin{figure}[t]
    \centering
    \subfigure[Single-column search scenario.]{
        \centering
        \includegraphics[width=0.225\textwidth]{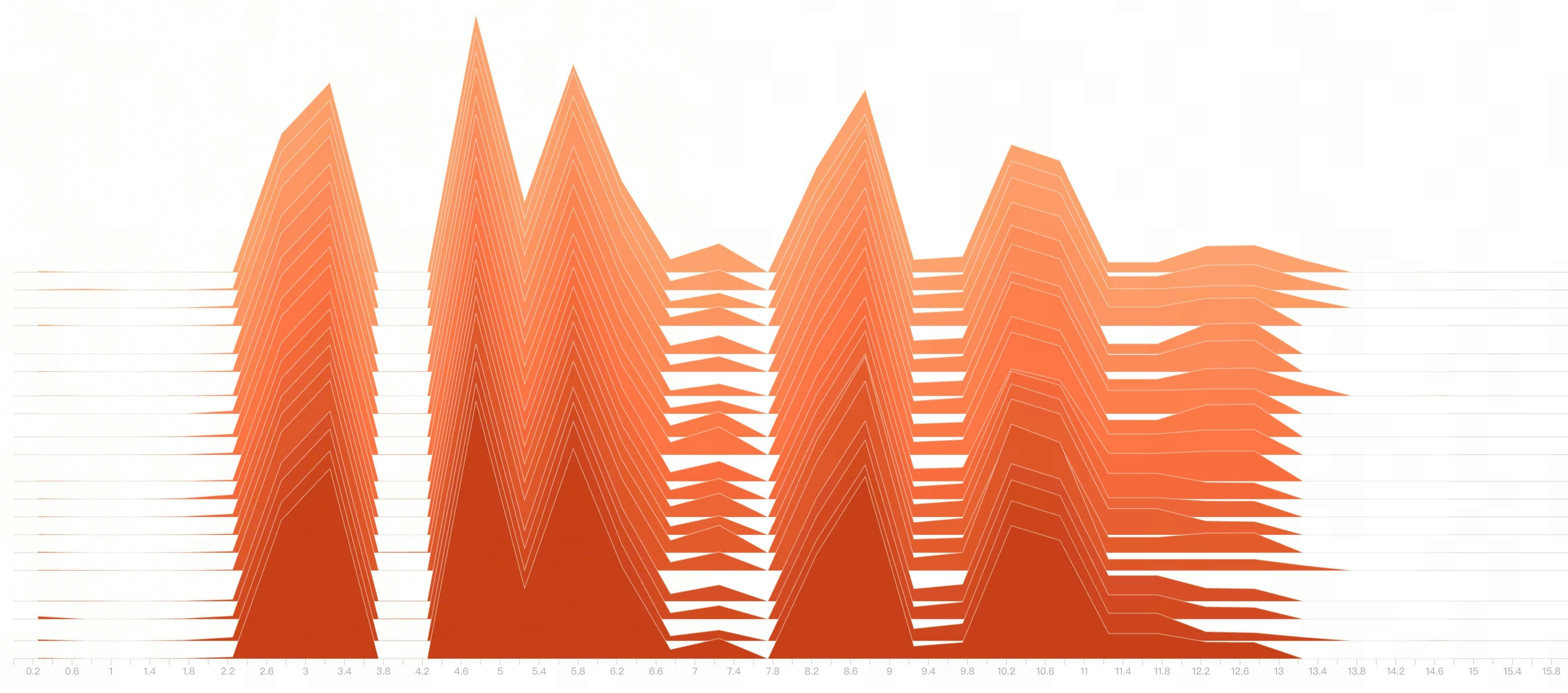}
    }
    \subfigure[Double-column search scenario.]{
        \centering
        \includegraphics[width=0.225\textwidth]{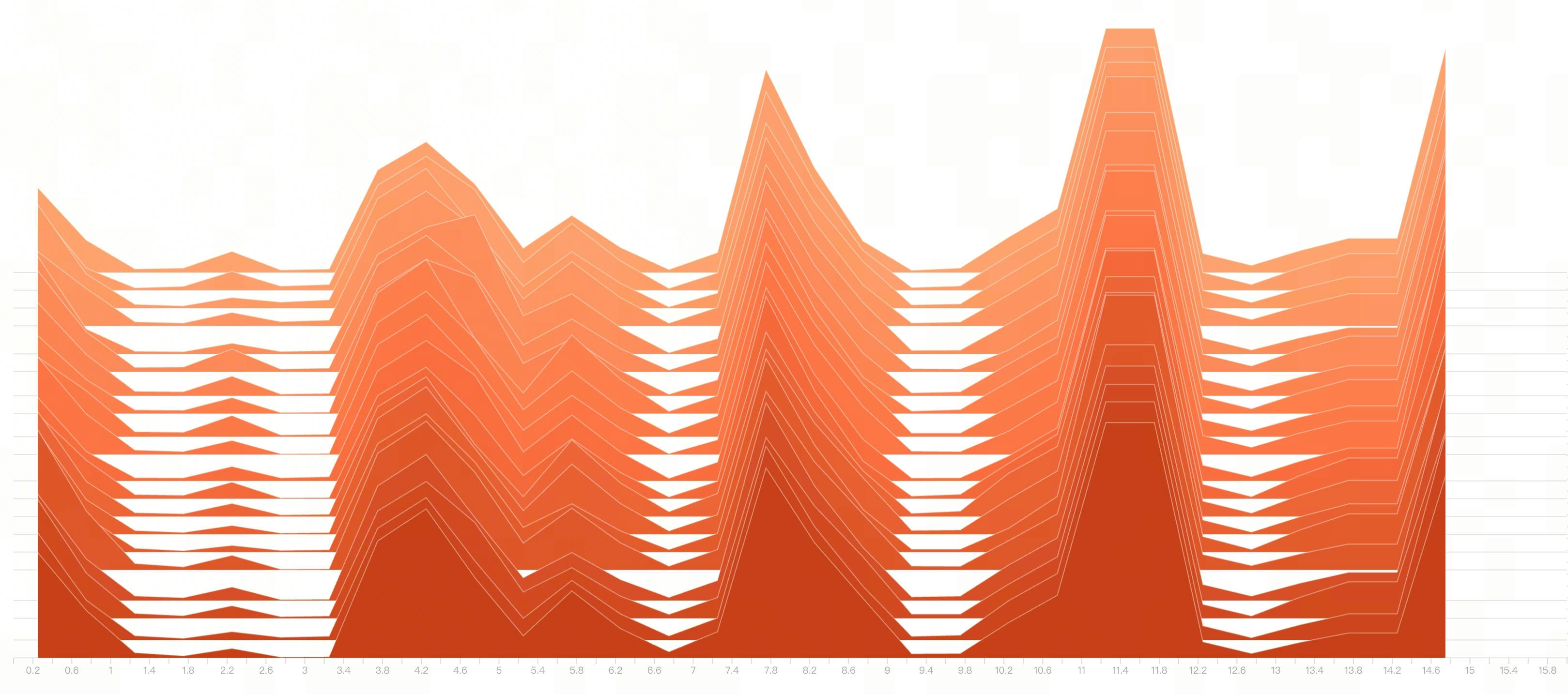}
    }
\caption{Attention distribution between scenario tokens and feature tokens. The x-axis represents the index of feature tokens and y-axis represents the average attention weights belong to the corresponding feature token.}
\label{fig:exp-vis2}
\end{figure}

\subsubsection{Visualizing the Attention Score}
A key component of the proposed approach MDL is the interaction between feature tokens and scenario/taks tokens.
Through the \emph{domain-aware attention}, the task tokens and scenario tokens can adaptively and selectively aggregate the rich information from feature tokens based on their own data characteristics for achieving differentiated modeling for multi-scenario and multi-task learning.
To better prove and understand the benefits brought by the domain-aware attention, we visualize the learned attention distributtion from different task tokens and scenario tokens in Figure~\ref{fig:exp-vis1} and Figure~\ref{fig:exp-vis2}.

From Figure~\ref{fig:exp-vis1}, we can observe that within the same layer, the attention distribution between different task tokens and feature tokens is different.
This indicates that different task tokens have learned distinct information for their respective distributions, thereby modeling the differences between tasks.
Moreover, for the attention distribution between the same task token and feature tokens, the attention distribution varies across different layers.
This indicates that as information interaction progresses layer by layer, task tokens can dynamically capture these changes and adaptively aggregate the corresponding information.
Similarly, from Figure~\ref{fig:exp-vis2}, we can observe that the the attention distribution between different scenario tokens and feature tokens is also different.

\begin{table}[t]
    \centering
    \caption{Results of online A/B tests on Douyin Search (relative improvements).}
    \label{tab:online}
        \begin{tabular}{ccc}
            \toprule
            Scenario & Change Query Rate & LT30 \\
            \midrule
            ALL & -0.3267\% & +0.0626\%  \\
            \cmidrule{1-3}
            Single-column Search & -0.2678\% & +0.0520\%  \\
            Double-column Search & -0.5079\% & +0.0674\%  \\
            Inner Search & -0.5492\% & +0.0630\%  \\
            \bottomrule
        \end{tabular}
\end{table}

\subsection{Online Experiments}
\label{exp:online}
To further verify the effectiveness of the proposed approach \modelname, we conducted an online A/B test on Douyin Search for one month.
The online base model is based on RankMixer~\cite{RankMixer}, with multi-scenario and multi-task learning implemented via MMoE~\cite{MMoE}.
The core online metrics include change query rate and LT30, as introduced in Section~\ref{subsec-evluation}.
Table~\ref{tab:online} reports the relative improvements in the different scenarios and the overall performance.
The proposed method yielded +0.0626\% improvement in LT30 and -0.3267\% reduction in change query rate.
As for each scenario, it also achieved significant improvements on the core metrics.

\section{Related Work}
\label{sec-rel}
In this section, we review the related works in two aspects, namely large-scale recommendation methods and multi-scenario multi-task recommendation methods.

\paratitle{Large-scale Recommendations.}
Inspired by the success of scaling laws in large language models (LLMs)~\cite{LLM1,LLM2}, the scaling laws of recommendation models have recently been widely studied and have achieved significant success in real-world industrial recommendation systems~\cite{wukong,RankMixer,KGBRank,HSTU,MTGR,OneTrans,hyformer}.
Typically, these methods increase the model parameter size and computational complexity by stacking various efficient feature interaction structures, thereby improving the recommendation performance.
One type of method focuses on scaling the parameters and complexity of the non-sequential feature interaction module for better performance.
For example, Wukong~\cite{wukong} scales the feature interaction module by introducing a stacked factorization machine-based architecture.
Rankmixer~\cite{RankMixer} introduces multi-head token mixing and pertoken feed-forward networks to scale recommender models, while maintaining inference latency and achieving superior performance.
FAT~\cite{KGBRank} scales recommendation models by introducing field-aware transformer with field-based interaction priors in attention.
Another typical method adopts the transformer-based structure for modeling non-sequential and sequential feature in a unified way and scaling constantly.
For instance, HSTU~\cite{HSTU} sequentializes the heterogeneous feature space and then performs sequence transduction conditioned on contextual and candidate signals.
MTGR~\cite{MTGR} further scales recommendation models based on HSTU  for acceleration and leakage avoidance.
OneTrans~\cite{OneTrans} introduces a unified transformer-based backbone that simultaneously models user-behavior sequences and feature interactions via a unified tokenizer, enabling efficient parameter scaling and superior performance.
Compared to these methods, our work further explores how large-scale recommendation models can better perform in multi-scenario and multi-task settings.

\paratitle{Multi-scenario and Multi-task Recommendations.}
In real-world industrial recommendation systems, a recommendation model usually needs to serve multiple scenarios and tasks.
Multi-scenario learning (MSL) and multi-task learning (MTL) in recommender system have been widely studied in previous works~\cite{sharedbottom,MMoE,Cross-Stitch,snr,PLE,HMoE,STAR,PEPNet,hinet,mmfi,yrr,MTL1}.
For MTL, most existing methods adopt a shared-specific architecture in which task-shared and task-specific components are combined together (\eg Sharedbottom~\cite{sharedbottom}, Cross-stitch~\cite{Cross-Stitch}, SNR~\cite{snr}).
In addition, Mixture-of-Experts (MoE) based structures have also been widely proposed, aiming to enhance the effectiveness of multi-task learning by designing various expert mechanisms (\eg MMoE~\cite{MMoE}, PLE~\cite{PLE}).
For MSL, it aims to use a unified model to learn data from multiple scenarios, thereby improving recommendation performance.
Most existing work also adopts a shared-specific architecture in which scenario-shared and scenario-specific components are combined together.
The framework used in MTL can also be applied in MSL by treating the predictions for different scenarios as distinct prediction tasks (\eg SharedBottom~\cite{sharedbottom}, MMoE~\cite{MMoE}).
In addition, some specialized multi-scenario methods have also been proposed~\cite{HMoE,STAR,PEPNet,hinet,mmfi,hc2,M2M,AdaSparse,SASS,TreeMS,SARNet,SAML}.
For example, HMoE~\cite{HMoE} proposes the hybrid of implicit and explicit MoE structure which learns the scenario and task relationships.
STAR~\cite{STAR} introduces a start topology structure where a centered network is used for all scenarios and scenario-specific network is generated for each scenario.
PEPNet~\cite{PEPNet} takes the scenario related prior information to adjust the feature embeddings and hidden units through gate mechanisms.
In contrast to these approaches which are based on the shared-specific architecture, we propose a novel multi-distribution learning framework from the perspective of tokenization, aiming to enhance the learning capability and unified modeling ability of large-scale recommendation models in multi-scenario and multi-task settings.

\section{Conclusion}
\label{sec-con}
In this paper, we propose a novel multi-distribution learning framework for multi-scenario learning (MSL) and multi-task learning (MTL) in large-scale industrial recommender system, named \textbf{MDL}.
Unlike existing MSL and MTL recommendation methods that are based on a \emph{shared-specific} architecture, the proposed approach inspired by large language models (LLMs) is to leverage prompt-based tokenization to achieve model predictions in various scenarios and tasks settings.
Specifically, MDL tokenizes feature, scenario, and task information into a unified tokenized format, and elobrates three key interaction mechanisms among these tokens.
Extensive experiments on both offline evaluation and online test have demonstrated the effectiveness.
The proposed approach MDL has been fully deployed on Douyin Search platform.

\balance
\bibliographystyle{ACM-Reference-Format}
\bibliography{main}

\end{document}